\documentclass[12 pt, amsfonts, amssymb,color]{article}

\usepackage{amsmath,amssymb,amsfonts,latexsym}
\usepackage{color} 

\usepackage{amsthm} 

\DeclareMathOperator{\sign}{sign}

\usepackage[all]{xy}
\def\R{\mathbb{R}}

\usepackage
{hyperref}

\begin{document}

\renewcommand\theequation{\arabic{section}.\arabic{equation}}
\catcode`@=11 \@addtoreset{equation}{section}
\newtheorem{defn}{Definition}[section]
\newtheorem{theorem}{Theorem}[section]
\newtheorem{example}{Example}[section]
\newtheorem{lem}{Lemma}[section]
\newtheorem{prop}{Proposition}[section]
\newtheorem{cor}{Corollary}[section]
\newcommand{\be}{\begin{equation}}
\newcommand{\ee}{\end{equation}}

\def\forms{{\textstyle\bigwedge}}
\newcommand{\equal}{\!\!\!&=&\!\!\!}
\newcommand{\rd}{\partial}
\newcommand{\g}{\hat {\cal G}}
\newcommand{\bo}{\bigodot}
\newcommand{\res}{\mathop{\mbox{\rm res}}}
\newcommand{\diag}{\mathop{\mbox{\rm diag}}}
\newcommand{\Tr}{\mathop{\mbox{\rm Tr}}}
\newcommand{\const}{\mbox{\rm const.}\;}
\newcommand{\cA}{{\cal A}}
\newcommand{\bA}{{\bf A}}
\newcommand{\Abar}{{\bar{A}}}
\newcommand{\cAbar}{{\bar{\cA}}}
\newcommand{\bAbar}{{\bar{\bA}}}
\newcommand{\cB}{{\cal B}}
\newcommand{\bB}{{\bf B}}
\newcommand{\Bbar}{{\bar{B}}}
\newcommand{\cBbar}{{\bar{\cB}}}
\newcommand{\bBbar}{{\bar{\bB}}}
\newcommand{\bC}{{\bf C}}
\newcommand{\cbar}{{\bar{c}}}
\newcommand{\Cbar}{{\bar{C}}}
\newcommand{\Hbar}{{\bar{H}}}
\newcommand{\cL}{\mathcal{L}}
\newcommand{\bL}{{\bf L}}
\newcommand{\Lbar}{{\bar{L}}}
\newcommand{\cLbar}{{\bar{\cL}}}
\newcommand{\bLbar}{{\bar{\bL}}}
\newcommand{\cM}{{\cal M}}
\newcommand{\bM}{{\bf M}}
\newcommand{\Mbar}{{\bar{M}}}
\newcommand{\cMbar}{{\bar{\cM}}}
\newcommand{\bMbar}{{\bar{\bM}}}
\newcommand{\cP}{{\cal P}}
\newcommand{\cQ}{{\cal Q}}
\newcommand{\bU}{{\bf U}}
\newcommand{\bR}{{\bf R}}
\newcommand{\cW}{{\cal W}}
\newcommand{\bW}{{\bf W}}
\newcommand{\bZ}{{\bf Z}}
\newcommand{\Wbar}{{\bar{W}}}
\newcommand{\Xbar}{{\bar{X}}}
\newcommand{\cWbar}{{\bar{\cW}}}
\newcommand{\bWbar}{{\bar{\bW}}}
\newcommand{\abar}{{\bar{a}}}
\newcommand{\nbar}{{\bar{n}}}
\newcommand{\pbar}{{\bar{p}}}
\newcommand{\tbar}{{\bar{t}}}
\newcommand{\ubar}{{\bar{u}}}
\newcommand{\utilde}{\tilde{u}}
\newcommand{\vbar}{{\bar{v}}}
\newcommand{\wbar}{{\bar{w}}}
\newcommand{\phibar}{{\bar{\phi}}}
\newcommand{\Psibar}{{\bar{\Psi}}}
\newcommand{\bLambda}{{\bf \Lambda}}
\newcommand{\bDelta}{{\bf \Delta}}
\newcommand{\p}{\partial}
\newcommand{\om}{{\Omega \cal G}}
\newcommand{\ID}{{\mathbb{D}}}

\def\ep{\epsilon}
\def\de{\delta}

\def\wt{\widetilde}
\def\fracpd#1#2{\frac{\partial #1}{\partial #2}}
\def\pd#1#2{\frac{\partial #1}{\partial #2}}

\def\<#1>{\langle#1\rangle}

 \def\rc#1{\begin{color}{red}#1\end{color}}
 \def\bc#1{\begin{color}{blue}#1\end{color}}


\title{Sundman transformation and alternative tangent  structures} 
\author{Jos\'e F. Cari\~nena$^1$\footnote{E-mail:jfc@unizar.es}, Eduardo Mart\'{\i}nez$^2$
\footnote{E-mail:emf@unizar.es} and Miguel C. Mu\~noz-Lecanda$^3$
\footnote{E-mail:miguel.carlos.munoz@upc.edu}\\
$^1$Departamento de F\'{\i}sica Te\'orica and IUMA, Universidad de Zaragoza,\\Pedro Cerbuna 12, 
E-50009 Zaragoza, Spain\\$^2$Departamento de    Matem\'atica Aplicada and 
IUMA, Universidad de Zaragoza,\\
Pedro Cerbuna 12, 
E-50009 Zaragoza, Spain\\
$^3$Departament de Matem\'atiques, Campus Nord U.P.C., Ed. C-3\\
C/ Jordi Girona 1. E-08034 Barcelona, Spain}

\date{ }

\maketitle

\begin{abstract}  A geometric approach to Sundman transformation defined by basic functions  for systems of second-order differential equations is developed and the necessity of a change of  the tangent structure by means of the function defining
the Sundman transformation is shown. Among other applications of such theory we study the linearisability of a system of second-order  differential equations and in particular the simplest case of a second-order  differential equation. The theory is illustrated with several examples.

\end{abstract}

\smallskip

	 
\bigskip

 
\bigskip


 
\section{Introduction} 

The  infinitesimal  time reparametrisation introduced by Levi-Civita \cite{LC04,LC06}, usually called  Sundman transformation  \cite{S13},   was used in \cite{LC20}   to   
 regularise the 2-dimensional Kepler problem and it has been proved to be 
 very efficient to deal  with many different problems in the theory of systems of differential equations and related physical problems.  Such a transformation, 
 at least from a geometric  perspective, 
is very intriguing because the time does not explicitly appear in the expressions for autonomous systems,   and  generalisations of 
Sundman transformation to deal with autonomous systems 
of first-order differential equations  can be used in the study of  linearisation of systems of differential equations  \cite{du94} and 
  in numerical solution of systems of differential equations (see e.g. \cite{BI,CHL,N76}).
These generalised Sundman transformations have also been used to solve many  interesting problems in classical mechanics  (see e.g. \cite{B11,BM07, 
GMT17,MCF17}) and  celestial mechanics \cite{BJ,B85,CIL,CCJM,CMS19}. 

The geometric approach  to the study of  autonomous systems of  first-order differential equations  has been very useful and its results 
have very much clarified many points involved in the theory. Moreover, as it is intrinsic and the results do not depend on a particular
 choice of coordinates, its methods may be generalised to infinite
dimensional systems (with some topological difficulties). So, such a system   is replaced 
by a vector field on a differentiable manifold $M$, and the set of its solutions provides the set of  the integral curves of the vector field   in a local coordinate system. 
The time plays the r\^ole 
of the parameter of the integral curves. It has recently been shown \cite{CMM21}  that a  Sundman  infinitesimal time 
reparametrisation should be understood as a change of the 
dynamical vector field, replacing it by a conformal one. This geometric interpretation was analysed in a more detailed way in \cite{CMM21}.

On the other hand,  the autonomous systems of second-order differential equations are also to be studied, not only for its own 
mathematical interest, but also because they play a crucial r\^ole 
 in classical mechanics and in the spectral problem in Quantum Mechanics via the time-independent Schr\"odinger equation.  Then, the usual way to proceed is to reduce the problem of a system of $n$ second-order differential equations to a system of $2n$ first-order differential equations,
 or in geometric terms, to relate such systems of second-order differential equations to a special kind of vector field $X$ on
  its tangent bundle $TM$,  one of the class of second-order differential equation vector fields, 
 hereafter shortened as SODE vector fields.  Then one can try to  consider the Sundman transformation in the framework of SODE vector fields   in a similar way, 
i.e. by  changing the SODE vector field $X$ describing 
 the  autonomous system to the vector field $f\, X$. The point however is 
that   unless $f=1$, $f\, X$ is not a SODE vector field anymore, and a more careful analysis is needed. Fortunately, the existence of alternative tangent   structures will be useful to solve this problem (see e.g. \cite{CMS19}).
 
The aim of this paper is to clarify the meaning of an   infinitesimal time reparametrisation for systems of second-order differential equations  from a geometric viewpoint. Section 2 is devoted to recall 
a generalisation of the classical Sundman transformation for systems of first-order differential equations
  from a geometric perspective developed in \cite{CMM21}, while Section 3 
starts by summarising  the main geometric  ingredients of the theory of systems of second-order differential equations and the symplectic approach to regular 
Lagrangian systems, as some preliminary ideas to show the importance of the tangent  structures  as well as  the possibility of alternative tangent structures,  and afterward an interesting example of alternative tangent structure 
  to be used  for geometrically understanding  Sundman transformation for 
systems of second-order differential equations
 is given. Linear  systems of second-order differential equations are reviewed in Section 4 from a geometric perspective. The explicit meaning of  Sundman
  transformation for  systems of second-order differential equations is introduced  in Section 5, and the simpler case of one-dimensional problems and the
   linearisability of second-order differential equations   is  developed in Section 6, while illustrative examples are collected in Section 7. Some final remarks and comments on future work are given in the last section.
 
\section{A geometric approach to generalised Sundman transformation} 

We first review the geometric generalisation of Sundman transformations given in \cite{CMM21}.  The classical Sundman transformation \cite{S13} is 
an infinitesimal scaling of time from  $t$ to a new fictitious time $\tau$ given by
 \begin{equation}   
    dt= r\, d\tau \Longleftrightarrow \frac{d\tau}{dt}=\frac 1 r,  \label{Sundman}
     \end{equation}  but it was slightly generalised to $dt= c\,r^\alpha\, d\tau$, where $c\in\mathbb{R}$ and $\alpha$ is a positive constant \cite{N76}, or more generally to $dt= f(r)\, d\tau$ \cite{B85,FS,SB}.  
 
     We first consider the geometric generalisation of  Sundman  transformation for  systems
    of first-order differential equations given in  \cite{CMM21}.  Recall 
that, as indicated in \cite{PRV}, we can consider an autonomous system of first-order differential equations 
\begin{equation}
   \frac{dx^i}{dt}=X^i(x^1,\ldots,x^n), \quad i=1,\ldots,n,\label{autsyst}
   \end{equation}
   and under   the generalisation of Sundman transformation defined by 
   \begin{equation}
    dt=f(x^1,\ldots,x^n)\ d\tau,  \quad f(x^1,\ldots,x^n)>0, \label{genSundman}
   \end{equation}
   it formally becomes 
   \begin{equation}
   \frac{dx^i}{d\tau}=f(x^1,\ldots,x^n)\ X^i(x^1,\ldots,x^n),\quad i=1,\ldots,n.\label{newautsys}
   \end{equation}
   
 In the geometric approach the system (\ref{autsyst}) has associated the vector field  $X$  with coordinate expression 
    \begin{equation}
    X= \sum_{i=1}^nX^i(x^1,\ldots,x^n)\pd{}{x^i},
       \end{equation}
  and  the solutions of the system of equations (\ref{autsyst}) provide us the integral curves of the vector field $X$. It has been proved in \cite{CMM21} that  if 
the curve  $\gamma(t)$   is an integral curve of $X$,
and we carry out the reparametrisation for which  the new parameter  $\tau$ is defined by the relation 
\begin{equation} 
\frac{d\tau}{dt}=\frac 1{ f(\gamma(t))},\label{genSundman2}
\end{equation} 
then  the reparametrised curve $\bar\gamma(\tau)$ such that  $\bar\gamma(\tau(t))=\gamma(t)$   is an integral curve of the vector field $f\, X$, 
whose integral curves are solutions of (\ref{newautsys}).

Remark that when each one of  the integral curves of a vector field $X$ is arbitrarily reparametrised  we obtain a new family of curves, which when 
they are    integral curves of a vector field $Y$,
  as the two vector fields $X$ and  $Y$ would  have the same local constants of motion, they generate the same 1-dimensional distribution and, at least locally, there
    exists a nonvanishing function $f$ such that $Y=f\, X$. We have checked that this is the case for  the reparametrisation defined by a Sundman transformation,  and then  $f$ is 
    the function defining such Sundman transformation. Therefore, from this geometric point of view the generalised Sundman transformation (\ref{genSundman}) corresponds
     to change the vector field $X$   by its conformally related one $\bar X=f\, X$.
   
It is noteworthy that the `velocity' with respect to the new time is different and so the new velocity $\bar v$ is  related to the old one by $\bar v^i=f\, v^i$, as a
 consequence of (\ref{genSundman}). 

The reinterpretation of this `infinitesimal time scaling' was used in \cite{CIL}  to deal with the   satellite theory described by Bond and Janin in \cite{BJ}.
 There are   many applications of these generalised Sundman transformations, and more details can be found in \cite{CMM21}.

\section{Alternative tangent structures and their applications}\label{atbs}

\subsection{Tangent structures and Lagrangian systems}
The geometric  formulation of Lagrangian Classical Mechanics makes use of 
the tangent  structure of the velocity phase space \cite{Cr81,Cr83a,dFLMV89,CIMM}.

Recall that given a $n$-dimensional manifold $Q$, its tangent bundle $TQ=\bigcup _{m\in Q}T_mQ$
can be endowed with a vector bundle structure $\tau_Q:TQ\to Q$, the fibres $\tau_Q^{-1}(m)=T_mQ$ being $n$-dimensional real linear spaces.
The  usual coordinate charts on the tangent bundle $TQ$ are induced from a chart on its base manifold $Q$. Given a coordinate chart $(U,\varphi)$ of $Q$, we can induce
  a chart on $\mathcal{U}=\tau_Q^{-1}(U) $ by the tangent map $\phi=T\varphi$, i.e. $\phi(m,v)=(\varphi(m),\varphi_{*m}(v))$. In other words, if $\pi^i:\mathbb{R}^n\to \mathbb{R}$, 
  $i=1,\ldots,n$,  are the natural projections on each factor and $q^i=\pi^i\circ \varphi$ are the coordinate functions, i.e.  $ \varphi=(q^1,\ldots,q^n)$, then  we can
    consider on $TQ$ the coordinate basis of the linear space of vector fields $\mathfrak{X}(U)$ usually denoted $\{\partial/\partial q^j\mid j=1,\ldots ,n\}$ and the corresponding dual basis for the linear space of 1-forms  $\Omega^1(U)$, $\{dq^j\mid j=1,\ldots ,n\}$. 
  Then a  vector $v$ on a point $q\in U$  is $v=v^j\,(\partial/\partial q^j)_{|q}$ and a covector $\zeta$ on such a point  is $\zeta=p_j\,(dq^j)_{|q}$, with  $v^j=\<dq^j,v>$ and $p_j=\<\zeta,\partial/\partial q^j>$ being the usual velocities and momenta. In this way each chart $(U,\varphi)$ of 
$Q$ provides us a trivialization of the tangent bundle on $\tau_Q^{-1}(U)\approx U\times \mathbb{R}^n$ and another one  on $\pi_Q^{-1}(U)\approx U\times \mathbb{R}^n$ of the 
  cotangent bundle $\pi_Q:T^*Q\to Q$.
   
   The tangent bundle $TQ$, as any other vector bundle on $Q$, has associated a vector field $\Delta$, usually called  Liouville vector field, generator of dilations along the fibres.
   Its local expression in the above mentioned  tangent bundle coordinates is 
   \begin{equation}
   \Delta(q,v)= \sum_{i=1}^nv^i\pd{}{v^i},\label{Lvf}
   \end{equation}
   and there is also a natural (1,1)-tensor field, usually  called vertical endomorphism, or simply tangent structure, which satisfies $\text{Im}\, S=\ker S$ and an integrability condition (see later on). 
   Its local coordinate expression in the usual tangent bundle coordinates  is
   \begin{equation}
   S= \sum_{i=1}^n\pd{}{v^i}\otimes dq^i.\label{defS}
   \end{equation}
  The base manifold  $Q$ can be identified to the zero section for $\tau_Q$, while vertical vector fields on the tangent bundle $TQ$ are those which are tangent to the fibres, and therefore they
   have  local expressions 
  \begin{equation}
  D(q,v)= \sum_{i=1}^nf^i(q,v)\pd{}{v^i}, \label{levvf}
  \end{equation}
  i.e. the vertical vectors are those of $\ker S$.
   In particular, the Liouville vector field $\Delta\in\mathfrak{X}(TQ)$ is a vertical vector field, i.e. $S(\Delta)=0$,  vanishing on the zero section and,  as the vertical endomorphism    $S$ is homogeneous of degree minus one in velocities,  such that
   \begin{equation}
   \mathcal{L}_\Delta S=-S.\label{LDS} 
   \end{equation} 
 Moreover,  if $D\in\mathfrak{X}(TQ)$ is the  vertical vector field given 
by  (\ref{levvf}) on $TQ$,  then $\mathcal{L}_D S=-S$ if and only if   $D(q,v)=\Delta(q,v)+ c^i(q)\partial/\partial {v^i}$, i.e. if $D$ differs from $\Delta$ in the 
 vertical lift of a vector field on the  base, 
 because 
 $$ 
 \mathcal{L}_D S= \sum_{i=1}^n\left[ D, \pd{}{v^i}\right]\otimes   dq^i=- \sum_{i,j=1}^n\pd {f^j}{v^i} \pd{}{v^j}\otimes     dq^i,
 $$
 and therefore $\mathcal{L}_D S=-S$ if and only if 
 $$f^i(q,v)=v^i+ c^i(q), \quad  i=1,\ldots,n.
 $$
 Note that this shows that $\Delta $ is the only vertical vector field on 
$TQ$ such that $\mathcal{L}_\Delta S=-S$ and vanishes on the zero section, i.e. $\Delta(q,0)=0$.

A special kind of vector fields on the tangent bundle $TQ$ is that of the 
so called second-order differential equation vector fields, to be shortened as SODE vector fields.
These vector fields $\Gamma$  are characterized by the property $S( \Gamma)=\Delta$, and the local coordinate expression of such a SODE vector field is
$$\Gamma(q,v)= \sum_{i=1}^n\left(v^i\pd{}{q^i}+f^i(q,v)\pd{}{v^i}\right).
$$
The SODE name for such vector fields  is due to the property that their integral curves  are solutions of the system 
\begin{equation}\left\{
\begin{array}{rcl} \dfrac{dq^i}{dt}&=&v^i\\\dfrac{dv^i}{dt}&=&f^i(q,v)\end{array}
\right. \quad  i=1,\ldots,n,
\end{equation}
and therefore the projections of such curves on the base manifold are solutions of the system of  second-order differential equations $\ddot q^i=f^i(q,\dot q)$, for $i=1,\ldots,n$.
 
 The Lagrangian  formalism is introduced as follows: Given a function $L\in C^\infty(TQ)$, we can define a 1-form $\theta_L$ on $TQ$ by $\theta_L=dL\circ S$ and the energy function $E_L\in C^\infty(TQ)$ by means of $E_L=\Delta(L)-L$.
Their local coordinate expressions are, respectively, 
\begin{equation}
\theta_L= \sum_{i=1}^n\pd L{v^i}\, dq^i,\qquad E_L= \sum_{i=1}^nv^i\pd L{v^i}-L.\label{tLEL}
\end{equation}

The Lagrangian function $L\in C^\infty(TQ)$  is said to be regular if $\omega_L=-d\theta_L$ is nondegenerate, i.e. $(\omega_L)^{\wedge n}\ne 0$. In this case we can
 define a Hamiltonian dynamical system on $TQ$ by $(TQ,\omega_L,E_L)$ and then the dynamical vector field $\Gamma$, which is a SODE vector field, is the solution of the
 dynamical equation 
 \begin{equation}
 i(\Gamma)\omega_L=dE_L.\label{dynL}
 \end{equation}

 Remark also that if $\Gamma$ is a SODE vector field the equation   (\ref{dynL}) is equivalent to 
 \begin{equation} \mathcal{L}_\Gamma\theta_L=dL,\label{dynL2}
 \end{equation}
  because  $ \mathcal{L}_\Gamma\theta_L= i(\Gamma)d\theta_L+d( i(\Gamma)\theta_L)$,  and when $\Gamma$ is a SODE we have that $ i(\Gamma)\theta_L=\Delta L$.
  
More details of the geometric approach to Lagrangian mechanics can be found in  \cite{Cr81,Cr83a,dFLMV89,CIMM}.

\subsection{Alternative tangent structures on a tangent bundle}

The use of alternative structures for the description of mechanical systems has been shown to be very useful for a better understanding of 
dynamics, providing unexpected results as, for instance,  recursion operators \cite{GM97,M78,MM84}  and non-Noether constants of motion \cite{CI}. 
The relevance and usefulness of the existence 
of alternative geometric structures have been shown in many recent works and have been summarized in a recent book \cite{CIMM}.

In particular, the tensorial   characterisation of linear structures and 
 vector bundle structures (partial linear structures) on a manifold have been analysed
in Chapter 3 of \cite{CIMM}. In summary, a {\sl linear structure} on a manifold 
$M$ is characterised (see  \cite{CIMM,IM12}) by the existence of a complete vector field $\Delta\in\mathfrak{X}(M)$ with only 
one non-degenerated critical point and such that $\mathcal{F}_\Delta^{(0)}=\mathbb{R}$ and  $\mathcal{F}_\Delta^{(1)}$ separates derivations, where $\mathcal{F}_\Delta^{(k)}$ 
is the set of functions on the manifold $M$ defined by 
$$\mathcal{F}_\Delta^{(k)}=\{f\in \mathcal{F}(M)\mid \Delta f=k\,f\}, 
\quad  k\in \{0\}\cup \mathbb{N}.$$ 
Recall that we say that $\mathcal{F}_\Delta^{(1)}$ separates derivations when given two different derivations $D_1$ and $D_2$ there exists a function $f\in 
\mathcal{F}_\Delta^{(1)}$ such that $D_1f\ne D_2f$

This vector field $\Delta$ is called Liouville vector field.
Similarly, a given $n$-dimensional manifold $M$  is a {\sl vector bundle} when there 
exists a complete vector field $\Delta$, also called Liouville vector field, such that the
 set of points $m$ on $M$ such that $\Delta(m)=0$
is a $k$-dimensional submanifold $Q$, the set of functions $\mathcal{F}_\Delta^{(0)}$ is an Abelian algebra whose spectrum is $Q$, and the equation 
$\Delta f=f$ admits $n-k $ functionally independent fibrewise-linear functions (see \cite{IM12}). In this case $d\mathcal{F}_\Delta^{(0)}$  and  
$d\mathcal{F}_\Delta^{(1)}$ span the set of 1-forms $\Omega^1(M)$ as a 
$\mathcal{F}_\Delta^{(0)}$-module, or, in other words,  $\Delta$ defines, at least locally, a partial linear structure and  a submersion $\pi:M\to Q$. It is possible to choose local coordinates $(x^1,\ldots,x^k)$  on the base manifold and
$(y^1,\ldots,y^{n-k})$ on the fibres in such a way that  (see \cite{CIMM})
\begin{equation}
\Delta=\sum_{\alpha=1}^{n-k}y^\alpha\pd{}{y^\alpha}.\label{Deltavf}
\end{equation} 

The particular case we are interested in is that of a tangent  bundle structure. The concept of almost-tangent structure on a manifold $M$ was introduced in \cite{CB60}  and 
\cite{E62} as a (1,1)-tensor field $S$ on the manifold $M$ such that at each point $p\in M$ the kernel of the linear map $S_p:T_pM\to T_pM$ coincides with its image.
 It follows that $S^2=0$  and that $M$ must be even dimensional, $\dim M=2n$. It is a particular case of the more general definition given in \cite{H}, and its study received attention
during the nineteen-seventies \cite{H,BC,CG,G72,YD}. 

The  almost-tangent structure $S$  is said to be integrable if its Nijenhuis tensor $N_S$ vanishes.  We recall that the Nijenhuis 
tensor $N_T$  of a (1,1)-tensor field $T$  on the manifold $M$ is a (1,2)-tensor field given by 
\begin{equation}
N_T(X_1,X_2)=[T(X_1),T(X_2)]+T^2([X_1,X_2])- T([T(X_1),X_2])-T([X_1,T(X_2)]),\  \forall X_1,X_2\in\mathfrak{X}(M),\label{Nijent}
\end{equation}
and therefore, when $T=S$, as $S^2=0$, 
$$N_S(X_1,X_2)=[S(X_1),S(X_2)]-S([S(X_1),X_2])-S([X_1,S(X_2)]),\quad \forall X_1,X_2\in\mathfrak{X}(M),$$
and consequently,  $N_S=0$ if, and only if, for every vector field $X\in  \mathfrak{X}(M)$  \cite{Cr83b}, 
$$\mathcal{L}_{S(X)} S=-(\mathcal{L}_{X}S)\circ S,
$$
because, for each vector field $Y\in \mathfrak{X}(M)$
$$(\mathcal{L}_{S(X)} S)(Y)=[S(X),S(Y)] -S([S(X),Y]),
$$
and 
$$
(\mathcal{L}_{X}S)(S(Y))=\mathcal{L}_{X}(S(S(Y)))-S([X,S(Y)])=-S([X,S(Y)]).
$$
The remarkable point is that the $N_S=0$ condition implies that the vertical distribution  defined by $\ker S=\textrm{Im}\,S$ is involutive, because if $X,Y\in \ker S=\text{Im}\, S$, then there exist $\widetilde X, \widetilde Y \in \mathfrak{X}(M)$ such that 
$S(\widetilde X)=X$ and $S(\widetilde Y)=Y$, and then 
$$[X,Y]=[S(\widetilde X), S(\widetilde Y)]=S([S(\widetilde X),\widetilde Y])+S([\widetilde X,S(\widetilde Y)]),$$
i.e. $[X,Y]$ lies in the image of $S$, which coincides with $\ker S$,
and therefore the distribution  defined as $\ker S$ is involutive, and then  integrable in the Frobenius sense. Recall that Frobenius theorem establishes that a 
distribution $\mathcal{D}$ is integrable if, and only if, it is involutive, i.e. $[\mathcal{D},\mathcal{D}]\subset \mathcal{D}$. In this case the manifold splits in leaves which are the integral submanifolds of $\mathcal{D}$ and integrable distributions are called foliations.

As pointed-out in \cite{Cr83b},
 at least  locally, the 
integral submanifolds of such foliation are the fibres of a fibration $\pi:M\to Q$  where $Q$ is a $n$-dimensional manifold. The vertical vector fields, which project on the zero vector field on $Q$,  are those taking values in $\ker S$. Then, if $X_1\in  \mathfrak{X}(M)$ and  $X_2\in  \mathfrak{X}(M)$ are projectable vector fields on the same vector field  $\bar X\in  \mathfrak{X}(Q)$, then as $X_2-X_1$ is a vertical vector field, $S(X_1)=S(X_2)$,  that is,  $X_1-X_2\in \ker S$. Conseqently, 
 if $X\in  \mathfrak{X}(M)$ projects on a vector  field 
$\bar X\in  \mathfrak{X}(Q)$, then $S(X)$ only depends on $\bar X$.   

\begin{lem} \quad\par{\it i)}  If $X\in \ker S$ and the vector field $Y\in \mathfrak{X}(M)$ is projectable, then $[X,Y]\in\ker S$.

ii) If  $X_1, X_2\in  \mathfrak{X}(M)$   are projectable  vector fields, then $[S(X_1),S(X_2)]=0$.
\end{lem}

{\sl Proof.-} {\it i)} As  the vector field $X$ is projectable, because it is  in $\ker S$,  and $Y\in \mathfrak{X}(M)$ is   projectable on  $\bar Y\in  \mathfrak{X}(Q)$,  then the vector field $[X,Y]$ is projectable on  $[0, \bar Y]=0$, i.e. 
$[X,Y]\in \ker S$.

{\it ii)}  The vanishing condition of the Nijenhuis tensor, $N_S(X_1,X_2)=0$, together with the result {\it i)} imply  that $[S(X_1),S(X_2)]=0$, because $[S(X_1),X_2])$ and $[X_1,S(X_2)])$ are in $\ker S$.

 $\hfill\Box$

If $(x^1,\ldots,x^n)$ are local coordinates on $Q$   we can choose local coordinates $(x^i,u^i)$ on $M$, and it was proved in \cite{Cr83b} that such  local coordinates $u^i$
can be  chosen  in such a way that  if $X_1,\ldots,X_n$, are vector fields on $M$
 projecting  on $\partial/\partial x^1,\ldots, \partial/\partial x^n$, respectively, as $[S(X_i),S(X_j)]=0$, the  local  coordinates on the fibres satisfy  $S(X_i)=\partial/\partial u^i$.
 The local   expression of $S$   is then $S=(\partial/\partial u^i)\otimes dx^i$.
    
 Note however that the
  coordinates $u^i$ are determined in this procedure only up to an additive constant on each fibre, i.e. they depend on a choice that plays the r\^ole of  zero section, because they are solutions of 
  the system  of differential equations $S(X_i)u^j=\delta_i^j, \  i,j=1\ldots n$,  and we can change $u^i$ by $\bar u^i=u^i+f^i(x)$. The ambiguity functions  $f^i$ are fixed by the choice of the zero section.
  Such a choice uniquely determines, for each coordinate system $(x^i)$ on the base manifold $Q$,  a system of local coordinates $(x^i,u^i)$ on $M$ such that the integrable almost tangent structure
  is given by  
  \begin{equation}
  S=\sum_{i=1}^n\pd{}{u^i}\otimes dx^i. \label{genS} 
  \end{equation}
  
  Given  a vector field on the base, $\bar X\in  \mathfrak{X}(Q)$, there is one vector field  $X\in \mathfrak{X}(M)$ such that $X$ and $\bar X$  are $\pi$-related, the relation $\mathcal{L}_{X}S=0$ holds, 
  and $X$ is tangent to the image of the zero section. In fact,  if 
\begin{equation}
\bar X(x)=\sum_{i=1}^n f^i(x)\pd{}{x^i},\label{bvf}
\end{equation} 
the coordinate expression of a vector field $X$   $\pi$-related to $\bar 
X$ is 
$$
 X(x,u)=\sum_{i=1}^n\left( f^i(x)\pd{}{x^i}+g^i (x,u)\pd{}{u^i}\right),
$$
and the condition  $\mathcal{L}_{X}S=0$ implies that
$$\pd{g^i}{u^j}-\pd {f^i}{x^j}=0, \quad i,j=1\ldots n \Longleftrightarrow g^i=\sum_{j=1}^n\pd {f^i}{x^j}u^j+c^i(x), \quad i=1\ldots n, 
$$
and then  the tangency condition implies $c^i(x)\equiv 0$, and hence the explicit coordinate expression of such a vector field $X$ is:
\begin{equation}
 X(x,u)=\sum_{i=1}^n f^i(x)\pd{}{x^i}+\sum_{i,j=1}^n\pd {f^i}{x^j}u^j\pd{}{u^i},\label{genclift}
\end{equation}
which formally looks like the complete lift of $\bar X$ from $Q$ to $M$.

Furthermore, there is a   uniquely-defined vertical vector field  $\Delta$ 
such that  $\mathcal{L}_\Delta S(X)=-S(X)$  for such vector fields (\ref{genclift}) and $\Delta=0$ on the zero section, because 
 $S(X)$ is then the vertical vector field
 $S(X)={\displaystyle \sum_{i=1}^n}f^i(x) (\partial /\partial {u^i})$, and if the local expression of the vertical field $\Delta$ is $\Delta={\displaystyle \sum_{i=1}^n}g^i(x,u) (\partial /\partial {u^i})$, as
 $$\left[\sum_{i=1}^nf^i(x) \pd{}{u^i},\sum_{j=1}^n g^j(x,u)\,\pd{}{u^j} \right]= \sum_{i,k=1}^nf^i(x)  \pd{g^k}{u^i} \pd{}{u^k},
 $$
  and $\mathcal{L}_\Delta S(X)=-S(X)$, it implies  $\partial g^k/\partial {u^i}=\delta^k_i, \  i,k=1\ldots n$,
  i.e. $g^k(x,u)=u^k+ c^k(x)$, while the vanishing condition on the zero section fixes $c^i(x)\equiv 0$, and hence  the explicit expression of $\Delta$ 
  is
  \begin{equation}
  \Delta = \sum_{i=1}^nu^i\pd{}{u^i}.\label{genDelta}
  \end{equation} 
 The vector field $\Delta$ provides a linear space structure to every fibre of $\pi:M\to Q$.
 
 The vector fields associated to systems of second-order differential equations with respect to the new integrable almost tangent structure given by  (\ref{genS}) are those of the form
 $$D(x,u)=\sum_{i=1}^n\left( u^i\pd{}{x^i}+ f^i(x,u)\pd{}{u^i}\right),
 $$
 and are characterized by $S(D)=\Delta$.

The r\^ole of the integrable almost tangent structure was clarified in \cite{CT} where it was clearly established 
that the almost tangent structure is responsible only of the affine structure of the tangent bundle  rather than of  its linear structure. It is the vector field $\Delta$ which selects the
 linear structure as indicated in the preceding paragraph.
 
 \subsection{An important example}
 
 A particularly important example of integrable almost tangent structure is the above mentioned one   of the tangent bundle $\tau_Q:TQ\to Q$, where $S$ is the vertical endomorphism (see e.g. \cite{Cr83a}). Alternative tangent structures have also been exhibited in \cite{dFLMV89}, but we are going to fix our attention on the following example which is a particular case of other more general method of construction,  given a vector field $X\in\mathfrak{X}(TQ)$ on a tangent bundle $\tau:TQ\to Q$, of an alternative tangent structure  such that the vector field $X$ is a SODE vector field with respect to the new tangent bundle structure.   

 Now,  given a constant sign basic function $\tau_Q^*h$  on $TQ$, with   $h\in C^\infty(Q)$, we can introduce a new  integrable almost tangent structure $(\tau_Q^*h)\, S$ on $TQ$, 
 simply denoted $\bar S=h\, S$, because $\ker (h\, S)=\textrm{Im}(h\, S)$. The vertical distribution defined by  $\ker (h\, S)$ coincides with the usual one defined by  $\ker S$ 
 and the set of leaves can be identified with the base manifold $Q$. If we choose a local set of coordinates for $Q$, $(q^1,\ldots,q^n)$,  and if $X_1,\ldots,X_n$, are 
 $\tau_Q$-projectable  vector fields  on $TQ$,
 $\tau_Q$-related to $\partial/\partial q^1, \ldots, \partial/\partial q^n$, respectively,  then the vector fields $h\,S(X_1), \ldots,h\,S(X_n)  $ 
  are pairwise commuting (by Lemma 3) 
  and there exist local coordinates $\bar v^1,\ldots,\bar  v^n$,  such that $S(h\,X_i)=h\,S(X_i)=\partial/\partial {\bar v}^i$, $i=1,\ldots, n$, and as $S(X_i)=\partial/\partial v^i$,  with the
   same choice for the  zero section, we can see that  $\bar S=(\tau_Q^*h)\, S$, has the local expression 
   $$\bar S=\sum_{i=1}^n\pd{}{\bar v^i}\otimes d{q^i}=h\,S,
   $$
   with the fibre coordinates being given by 
    $h\,{\bar v}^i=  v^i$, $i=1,\ldots, n$. 
     Of course the vector field $\bar \Delta$ coincides with the usual generator of dilations, the standard Liouville vector field $\Delta$. 
     
     Note that as the local expression of the new (1,1)-tensor $\bar S$ is similar to  that of $S$, the condition 
$N_{h\, S}=0$ follows from $N_S=0$.
     
     It is also to be remarked that the $\bar v^i$ are not coordinates adapted to the original  tangent bundle structure because the 1-forms $\alpha^i$ defining such coordinates of the vector $v$ 
      by $\bar v^i=\alpha^i(v)$ are not exact but $\alpha^i= (1/h ) dq^i$,  and  they correspond to the so called quasi-velocities (see \cite{heard,CNS, CGMS14b,CS21} and references therein). 
      
      In summary, going from $S$ to $\bar S=h\, S$ we obtain a new tangent structure on $TQ$, with its corresponding Liouville vector field $\bar\Delta=\Delta$ and a new concept of SODE vector field with respect to the new tangent structure.
   
\section{Linear  systems of second-order differential equations}
\label{section:linear.sodes}

Sundman transformation was first introduced to deal with systems of second-order differential equations, Newton equations of motion, and it was used to study the linearisation of 
  systems of second-order differential equations (see e.g. \cite{MM11,JM13}). We first recall some relevant concepts of the geometry of  autonomous
  systems of  second-order differential equations.  As mentioned in the preceding section, given such a system 
  \begin{equation}
  \frac{d^2x^i}{dt^2}=X^i(x^1,\ldots,x^n,\dot x^1,\ldots,\dot x^n),  \quad i=1,\ldots,n,\label{secautsyst}
   \end{equation}
   it has associated a system of first-order differential equations 
    \begin{equation}\left\{
    \begin{array}{rcl}
  {\displaystyle \frac{dx^i}{dt}}&=&v^i\\   {\displaystyle\frac{dv^i}{dt}}&=&X^i(x^1,\ldots,x^n,v^1,\ldots, v^n)
  \end{array}\right. \quad i=1,\ldots,n,\label{2ndautsyst}
   \end{equation}
whose solutions define the integral curves of the second-order differential equation vector field $\Gamma$  on $TQ$
 \begin{equation}
\Gamma= \sum_{i=1}^n\left(v^i\pd{}{x^i}+X^i(x,v)\pd{}{v^i}\right),\label{sodevf}
 \end{equation}
  i.e.  a system of second-order differential equations can be dealt with 
a SODE vector field $\Gamma$ on the tangent bundle.
 Such vector fields can be  characterised in several ways. For instance by being sections of the two structures of vector bundle 
 of $T(TQ)$ over $TQ$, $\tau_{TQ}:T(TQ)\to TQ$ and $T\tau_Q:T(TQ)\to TQ$,  or alternatively, as indicated in the preceding Section, by $S(\Gamma)=\Delta$. 
But, when considering an alternative tangent structure there will be different second-order differential 
 equations vector fields with respect to the new tangent structure.

An interesting particular case is  that of   systems of second-order differential equations that are linear in a given coordinate system, i.e. a tangent bundle chart induced  from a chart for the base manifold $Q$ on an open set $U$ of $Q$.  In this case the functions $X^i$ on $U$ appearing in (\ref{sodevf}) must be  of the form $X^i(x,v)= {\displaystyle \sum_{j=1}^n}\left(A^i\,_j \, x^j+B^i\,_j \, v^j\right)$, where $A^i\,_j $ and $B^i\,_j $ are real 
numbers. In geometric terms,  this property can be written as $[\tilde \Delta,\Gamma]=0$, with $\tilde \Delta$ being the vector field on $U$ such that $\tilde \Delta=\Delta_Q +\Delta $, where $\Delta_Q$ and $\Delta$ are the vector fields on $U$ with local expressions $\Delta_Q =  {\displaystyle \sum_{i=1}^n} x^i (\partial/\partial x^i)$ and $\Delta={\displaystyle \sum_{i=1}^n} v^i (\partial/\partial v^i)$, because 
\begin{equation}
  \begin{array}{rcl}[\tilde \Delta,\Gamma]&=&{\displaystyle\left[ \sum_{i=1}^n\left(x^i\pd{}{x^i}+v^i\pd{}{v^i}\right), \sum_{j=1}^n\left(v^j\pd{}{x^j}+X^j(x , v)\pd{}{v^j} \right)\right] }= \\
 &=& {\displaystyle\sum_{j, k=1}^n\left(x^j\pd{X^k}{x^j}\pd{}{v^k}+v^j\pd{X^k}{v^j}\pd{}{v^k}\right)-\sum_{j=1}^nX^j(x , v)\pd{}{v^j}},
  \end{array}\label{conmDeltaG}
 \end{equation}
 and therefore,    $ [\tilde \Delta,\Gamma]=0$, if,  and only if, each component $X^i$, for $=1,\ldots,n$, satisfies
 $$
\sum_{j=1}^n\left(x^j\pd{X^i}{x^j}+v^j\pd{X^i}{v^j}\right) =X^i(x , v),  \quad i=1,\ldots,n,
$$
which, according to Euler theorem of homogeneous functions,  mean that the components $X^i$ are homogeneous functions  of degree 1, 
and this implies (see e.g \cite{KMS}, p. 213) that there exist  real constants  $ A^i\,_j$ and  $ B^i\,_j$   such that  
\begin{equation}X^i(x,v)= \sum_{j=1}^n(A^i\,_j\,x^j+B^i\,_j \, v^j),\qquad \quad i=1,\ldots,n.\label{totlinear}
\end{equation} 
Remark that the vector field $\tilde \Delta$ is not intrinsic but it depends on the choice of the chart on  $U$. More explicitly, the vector field $\tilde \Delta$ is the complete
 lift of $\Delta_Q$, and hence this notion of linearity of a SODE depends on the chart that has been used. Note  also that  if $\Delta_Q$ is globally defined,  then so is $\tilde{\Delta}$.
Recall that the existence of such   globally defined complete vector field $\Delta_Q$ implies that $Q$ is a linear space.

\medskip

Another subset of interesting systems are those which are linear in velocities, i.e. characterised by functions $X^i(x,v)$ in (\ref{sodevf}) for which there exist real functions $B^i\,_j$  such that 
\begin{equation}X^i(x,v)= \sum_{j=1}^n B^i\,_j(x) \, v^j,\qquad \quad i=1,\ldots,n.\label{vlinear}
\end{equation}
Such fibre-linear systems can be characterised in terms of the partial linear structure
 of the tangent bundle $TQ$, which  allows us to define the Liouville vector field $\Delta$. In fact, we can see that a SODE  vector field $\Gamma $ is  linear (in velocities) if $\<dv^i, [\Delta ,X]>=0$, 
 for all indices $i=1,\ldots , n$, because if $\Gamma $  is a SODE  given by (\ref{sodevf}), then 
$$[\Delta,\Gamma]=\sum_{i=1}^nv^i\pd{}{x^i}+\sum_{j,k=1}^n\left(v^k\pd{X^j}{v^k}-X^j
\right)\pd{}{v^j},
$$
and therefore, as 
$$\<dv^i, [\Delta ,X]>=\sum_{k=1}^nv^k\pd{X^i}{v^k}-X^i,\qquad i=1,\ldots, n,
$$
we see that $\<dv^i, [\Delta ,X]>=0$, for $i=1,\ldots , n$,  if and only if there exist $n^2$ real basic functions $A^i\,_j(x)$ such that $X^i={\displaystyle \sum_{j=1}^n}A^i\,_j(x)\, v^j$, i.e. the second subsystem of the associated system of differential equations for the determination of the integral curves of $\Gamma$ is linear in velocities. 

\medskip

We may also be interested in the case of systems of inhomogeneous linear second-order differential equations, for which the components $X^i$ of the vector field $X$ 
in (\ref{2ndautsyst}) must be  of the form $X^i(x,v)= {\displaystyle\sum_{j=1}^n  \left(A^i\,_j \, x^j+B^i\,_j \, v^j\right)}+C^i$, with $A^i\,_j$, $B^i\,_j$ and $C^i$ real constants, i.e. the vector field $\Gamma$ is a sum of $\Gamma=\Gamma _0+\Gamma_{-1}$ 
where 
\begin{equation}
\Gamma_0= \sum_{i=1}^n\left(v^i\pd{}{x^i}+\sum_{j=1}^n(A^i\,_j \, x^j+B^i\,_j \, v^j)\pd{}{v^i}\right)\,,\quad
 \Gamma_{-1}= \sum_{i=1}^nC^i\pd{}{v^i},\label{desilevf}
 \end{equation}
which satisfy
$$[\tilde \Delta, \Gamma_0]=0,\qquad [\tilde \Delta, \Gamma_{-1}]=- \Gamma_{-1},
$$
and therefore, $[\tilde \Delta,\Gamma]=-\Gamma_{-1}$ what implies that $[\tilde \Delta,\Gamma]$ is a vertical vector field with constant coefficients such that 
$$[\tilde \Delta,[\tilde \Delta, \Gamma]]=-[\tilde \Delta, \Gamma].$$

This last property characterises the vector fields corresponding to inhomogeneous linear second-order differential equations in a given chart, because this shows that if the vector field 
$[\tilde \Delta, \Gamma]$ is a constant vertical field $ - {\displaystyle\sum_{i=1}^n} C^i \partial/\partial v^i$ and, moreover,  $[\tilde \Delta,[\tilde \Delta, \Gamma]]=-[\tilde \Delta, \Gamma]$, then (\ref{conmDeltaG}) shows that the components $X^i$ satisfy $\tilde \Delta (X^i)-X^i=-C^i$,
and therefore, there exist $2n^2+n$ real constants $A^i\,_j$, $B^i\,_j$ and   $C^i$ such that $X^i= {\displaystyle\sum_{j=1}^n  \left(A^i\,_j \, x^j+B^i\,_j \, v^j\right)}+C^i$.

\medskip

Our aim is to study under which circumstances a given systems of  second-order differential equations can be transformed by an appropriate Sundman transformation into a linear or linear in velocities system. The properties of a Sundman transformation of a  systems of  second-order differential equations are analysed in next Section.

Observe that we have characterised different kinds of SODE  systems that we can integrate. These characterisations are not intrinsic but depend on the existence of an appropriate 
chart.
\section{Sundman transformation for systems of  second-order differential equations}\label{sundsode}
  
  Let us analyse now the geometric approach to Sundman transformation for such systems. As a system of second-order differential equations is geometrically described by a SODE 
  vector field, $\Gamma$, which is of a special kind of vector fields in the tangent bundle $TQ$,    the theory developed for systems of first-order differential equations   suggests us
   to proceed by similarity and  obtain  the transformed vector field by multiplication with the function defining the Sundman transformation. However as now the second derivatives appear, 
   maybe this approach should be modified. On the other hand, when multiplying by the function $f$ the SODE characteristic property  is lost. This leads us to examine this definition more carefully,  
   and it will be shown that we can overcome these two problems in the particularly  important case of the function $f$ defining the Sundman transformation being  a constant sign
   basic function,  and this fact will be assumed hereafter without explicit mention. 
      
 Actually, when applying a Sundman transformation (\ref{genSundman}) to the system (\ref{2ndautsyst}),  as indicated above, the new velocities with respect to the new time,  $\bar v$, are related to the previous ones  by  $\bar v^i=f\, v^i$, as a
 consequence of (\ref{genSundman2}), because given a curve, 
 $$
v^{i}=\frac{dx^{i}}{d t}=\frac{d x^{i}}{d \tau}\,\frac{d \tau}{d t}=\frac{1}{f}\bar{v}^{i}\, ,
$$ and therefore, as  $f$ is a basic function,
$$
\frac{\partial}{\partial v^{i}}={\displaystyle\sum_{j=1}^n}\frac{\partial\bar{v}^{j}}{\partial v^{i}}\frac{\partial}{\partial \bar{v}^{j}}=f\frac{\partial}{\partial \bar{v}^{i}}\, .
$$
This suggests the use of non-natural coordinates in $TQ$, the so called quasi-velocities \cite{heard,CNS, CGMS14b,CS21},
which have been shown to be very useful, for instance, in the study of Chaplygin systems \cite{BBM15b}.  
In the case we are dealing with, it  amounts to consider $\{f\,dx^1,\ldots, f\,dx^n\}$ as a nonholonomic basis of the  
 $  C^\infty(\mathbb{R}^n)$-module of sections of  $\tau:T\mathbb{R}^n\to \mathbb{R}^n$ (see \cite{CNS,CS21} and references therein). The corresponding dual basis
 of the module of vector fields on $TQ$ is made up by  
 $\{Y_k=f^{-1}\,\partial/\partial{x^k}\mid k=1,\ldots,n\}$, and then,
as 
$$
[Y_i,Y_j]=\left [f^{-1}\,\pd {}{x^i},f^{-1}\,\pd {}{x^j}\right] =-f^{-3}\pd{f}{x^i}\,\pd {}{x^j}+ f^{-3}\pd{f}{x^j}\,\pd 
{}{x^i},
$$
i.e.
$$[Y_i,Y_j]=\ \frac 1{f^2} \left(\pd{f}{x^j}Y_i-\pd{f}{x^i}\,Y_j\right)=\sum_{k=1}^j\gamma_{ij}^k \,Y_k ,
$$
we have that $[Y_i,Y_j]={\displaystyle \sum_{k=1}^n}\gamma_{ij}^k\, Y_k$, 
where the Hamel symbol $\gamma_{ij}^k$, necessary to write the Hamel-Boltzman equations corresponding to the dynamical evolution in the Lagrangian 
formalism  \cite{heard,CGMS14b,CS21}, is given by 
$$\gamma_{ij}^k=\frac 1{f^2}  \left(\pd{f}{x^j}\, \delta^k_i- \pd{f}{x^i} \,\delta^k_j\right).$$

 With this in mind we have a coordinate change to a new coordinate system on the manifold $TQ$ 
$$
(x^{i},v^{i})\mapsto(\bar{x}^{i},\bar{v}^{i})\ ,\quad  \bar{x}^{i}=x^{i} ,  \,\,\, \bar{v}^{i}=fv^{i}\, ,
$$
where we will keep the notation $\bar{x}^{i}$ for clarity.
Observe
 that, by direct calculus, we obtain
$$
\left\{
\begin{array}{rcl}{\displaystyle
\pd{}{x^{j}}}&=&
{\displaystyle\pd{}{\bar{x}^{j}}+\sum_{i=1}^n\pd f{x^j}v^{i}\frac{\partial}{\partial \bar{v}^{i}}=\pd{}{\bar x^j}+
 \sum_{i=1}^n\frac 1f \pd{f}{\bar x^j}\bar v^i\pd{}{\bar v^i}}\\
{\displaystyle\pd{}{v^j}}&=&f{\displaystyle\pd{}{\bar v^j}} 
\end{array}
\right. .
$$
Correspondingly,
$$
\left\{
\begin{array}{rcl}
d\bar x^k&=&dx^k\\
d\bar v^k&=&{\displaystyle \sum_{l=1}^n\frac 1f \pd{f}{\bar x^l}\bar v^k}dx^l+f\,dv^k
\end{array}
\right. ,
$$
and therefore,
$$
\left\{
\begin{array}{rcl}
dx^k&=&d\bar x^k\\
dv^k&=&{\displaystyle \frac 1f\left( -\sum_{l=1}^n\frac 1f \pd{f}{\bar x^l}\bar v^k\, d\bar x^l+d\bar v^k\right)}
\end{array}
\right. .
$$

Consequently, the new coordinate expression of the vector field $\Gamma$ given by (\ref{sodevf})
is
\begin{equation}
\Gamma(x,\bar v)=\sum_{i=1}^n\frac{\bar v^i}f\pd{}{\bar x^i}+\sum_{i=1}^n\left(fX^i(x,\bar v/f)+\bar v^i\sum_{j=1}^n\frac 1{f^2} \pd{f}{\bar x^j}\bar v^j\right)\pd{}{\bar v^i}.
\label{sodevfqc}
\end{equation}

To analyse the behaviour  under the generalised Sundman transformation (\ref{genSundman}) of the system (\ref{secautsyst}), we can use the operational relations 
\begin{equation}
\frac {d}{dt}=\frac 1f\frac {d}{d\tau}, \qquad \frac{d^2}{dt^2}=\frac 1f\frac {d}{d\tau}\left(\left(\frac 1f\right)\frac {d}{d\tau} \right)=\frac 1{f^2} \frac{d^2}{d\tau^2}-\frac 1{f^3}\frac {df}{d\tau}\frac {d}{d\tau},\label{dtdtau}
\end{equation}
and hence, if $x(t)$ is a solution of (\ref{secautsyst}), and the new parameter $\tau$ is given by 
$$\tau(t)=\int^t\frac 1{f(\eta)}\, d\eta,
$$
then $\bar x(\tau)$ such that $\bar x(\tau(t))=x(t)$ satisfies the system of second-order differential equations
$$\frac{d^2\bar x}{d\tau^2}=\bar X^i\left(\bar x,\frac{d\bar x^i}{d\tau}\right) \quad i=1,\ldots, n,
$$
with 
\begin{equation}
\bar X^i\left(\bar x,\frac{d\bar x}{d\tau}\right) ={f^2}\, X^i\left (\bar x, \frac 1f \frac{d\bar x}{d\tau}\right)+\frac d{d\tau}(\log f)\, \frac{d\bar x^i}{d\tau},\label{newX}
\end{equation}    
together with the condition $\bar v^i=f\,v^i$, i.e. $d\bar x^i/d\tau=f\, dx^i/dt$.
The system can be rewritten as 
\begin{equation}
\left\{\begin{array}{rcl}
  \dfrac{d \bar x^i}{d\tau}&=&f\, v^i=\bar v^i\\
  \dfrac{d\bar v^i}{d\tau}&=&{f^2}\, X^i\left (\bar x, \dfrac 1f \,\bar v\right)+\dfrac d{d\tau}(\log f)\, \bar v^i
\end{array}\right..\label{systnewX}
\end{equation}

This shows that the images under the  generalised Sundman transformation of the integral curves of the vector field $\Gamma$ given by (\ref{sodevf}),       
are the integral curves of the vector field
$\bar{\Gamma}(\bar x,\bar v)$ given by               
\begin{equation}
\bar{\Gamma}(\bar x,\bar v)= {\displaystyle\sum_{j=1}^n\bar{v}^{j}\frac{\partial}{\partial \bar{x}^{j}}}+{\displaystyle\sum_{i=1}^n}
\left({\displaystyle\sum_{j=1}^n}\frac{1}{f}\bar{v}^{j}\frac{\partial f}{\partial \bar x^{j}}\bar{v}^{i}+f^{2}X^{i}\left (\bar x, \dfrac 1f \,\bar v\right)\right)
\frac{\partial}{\partial \bar{v}^{i}}, \label{barGama}
\end{equation}
and a simple comparison with (\ref{sodevfqc}) shows that 
\begin{equation}
\bar{\Gamma}(\bar x,\bar v)=f(\bar x) \Gamma(\bar x,\bar v),
\label{barGamafGama}
\end{equation}
and therefore we see that, under the action of the  Sundman transformation,  the  associated SODE vector field     
$\Gamma$ given by (\ref{sodevf}) is  multiplied by the  function  $f\in C^\infty(Q)$, $\bar\Gamma=f\, \Gamma$. Moreover,
remark that  the new vector field $\bar\Gamma=f\, \Gamma$  is not a SODE vector field anymore. However, in the very relevant special class of Sundman transformations with 
 $f$ being a constant sign basic function, $f\in C^\infty(Q)$,  it is clear that   the new vector field $\bar\Gamma=f\, \Gamma$
 is a   SODE vector field with respect to the new tangent structure $\bar 
S$ related to $S$ as $\bar S=f^{-1}\,S$. Therefore the Sundman transformation amounts to multiply the vector field $\Gamma$ by the function $f$ but also to consider a new tangent structure, and then the vector field $\bar{\Gamma}$  is now a SODE with respect to the  new tangent structure defined by 
$$
\bar{S}=f^{-1}S=f^{-1}\sum_{j=1}^n\left(\frac{\partial}{\partial v^{j}}\otimes d x^{j}\right)=\sum_{i=1}^n\frac{\partial}{\partial \bar{v}^{i}}\otimes d \bar{x}^{j}\, ,
$$
with Liouville vector field 
$$
\bar \Delta=\sum_{j=1}^n\bar{v}^{j}\frac{\partial}{\partial \bar{v}^{j}}=\sum_{j=1}^nv^{j}\frac{\partial}{\partial v^{j}}=\Delta\, .
$$
In summary, if $\Gamma$ is a SODE vector field for the usual tangent bundle structure, then $\bar \Gamma$ 
is  a SODE vector field for the tangent structure defined by $\bar S=(1/f)S$.

 It is then clear from the expression (\ref{barGama})  of the vector field $\bar\Gamma$ in quasi-coordinates  that, in the simpler and relevant  case in which the $X^i$ are basic  functions, 
 the new vector field is not linear in the  associated chart. Conversely, 
in a general case  one can try to determine, if possible,  the function $f$ in such a way that the vector field $\bar \Gamma$ be linear in the fibre coordinates of the corresponding associated system of quasi-coordinates, i.e. it is fibre-linear in the new  tangent structure.

 Remark also that if we apply  successively to the system (\ref{2ndautsyst})  two Sundman transformations (\ref{genSundman}), characterised respectively  by the functions 
 $f_1\in C^\infty(Q)$ and $f_2\in C^\infty(Q)$,   we obtain the system obtained by applying the Sundman transformation
 defined by the product  $f_2\, f_1\in C^\infty(Q)$, and that the transformed vector field is a SODE vector field with respect to the new tangent structure $\bar{\bar S} =f_2^{-1}\, f_1^{-1}\,S$.
 Moreover, as  the set of positive real functions is an Abelian multiplicative Lie group, the Sundman transformation corresponding to $f^{-1}$ is the inverse of the Sundman transformation defined by $f$.

According to the previous comments, given  an arbitrary SODE field $\Gamma$, it may exist a positive basic  function $f$ such that  the vector field $\bar \Gamma$,  which is  a SODE  vector field
    with respect to the    new almost tangent structure $\bar S=f^{-1} \,S$, be  linear, or linear in velocities, in the new tangent structure.   Note  that when we consider a Sundman transformation (\ref{genSundman}), characterised by the functions $f\in C^\infty(Q)$, not only  the given vector field      $\Gamma$ should be 
transformed to     $\bar\Gamma$, but we should change the tangent bundle structure $S$ to  $\bar S=f^{-1} \,S$. But then  there is also a new chart for $TQ$ determined by the quasi-coordinates $(x^i,\bar v^i)$. 
The question is when is  it possible  to choose the function $f$ in order 
to the transformed vector field $\bar \Gamma$ be fibre-linear with respect to  the changed tangent bundle structure. 
This happens when the function   $f$ is  such that $ [\tilde \Delta,\bar \Gamma]=0$. 
    Similarly, sometimes the function $f$ can  be chosen such that the vector field $\bar \Gamma $   corresponds to a system of inhomogeneous linear differential equations.   
    Finally, it is noteworthy that very often the existence of constants of motion can be used in such a way that the reduced system is linear, even if the original system is nonlinear, as it will be shown  by means of an illustrative example.

The time evolution in terms of the new time $\tau$ of the new quasi-coordinates obtained under the given Sundman transformation, according to the second equation in   (\ref{systnewX}) is
 \begin{equation}
 \frac{d\bar v^i}{d\tau}=  \sum_{j=1}^n f\pd f{x^j}v^j\,v^i+f^2\, X^i(x,v)=\sum_{j=1}^n \frac 1f \pd f{\bar x^j}\bar v^j\,\bar v^i+f^2\, X^i\left(\bar x,\frac{\bar v}f\right),\label{teSTsystem}
\end{equation}
 and therefore the projection on the base manifold of the integral curves 
of $\bar\Gamma$ are solutions of the system of second-order differential equations
  \begin{equation}
   \frac{d^2\bar x^i}{d\tau ^2}= \left(
{\displaystyle\sum_{j=1}^n}\frac{1}{f} \frac{\partial f}{\partial \bar x^{j}} \frac{d\bar x^j}{d\tau}\right) \frac{d\bar x^i}{d\tau}+f^{2}X^{i}\left(\bar x,\frac 1f \frac{d\bar x}{d\tau}\right) .\label{teSTsystem2}
 \end{equation}

\section{Linearization of scalar SODEs}

\subsection{An example in Classical Mechanics}

In this section we consider the   case of an autonomous  one-dimensional SODE, i.e. $n=1$.
We first consider, because of its usefulness in mechanics,   the simplest case of   $X^1=F$  being a basic function. Then,   from the expression (\ref{teSTsystem}) 
we obtain  the equation for the integral curves of the  vector field corresponding to $\Gamma=v\,\partial/\partial q+F(q)\, \partial/\partial v$:
  \begin{equation}
  \frac{d\bar v}{d\tau}=  \frac 1f  f'(q)\,\bar v\,\bar v+f^2\, F(q),\label{eqbarvFq}
 \end{equation}
 where $q$ is the local coordinate in the 1-dimensional manifold $Q$. 
 
In the particular case of a system defined by a potential function $\mathcal{V}$,  where $F=-\mathcal{V}'$, the corresponding SODE vector field $\Gamma$ in coordinates $(q,v)$ and the transformed vector field  $\bar{\Gamma}=f\, \Gamma$ in coordinates $(\bar{q}, \bar{v})$ are given, respectively, by 
$$
\Gamma(q,v)=v\frac{\partial}{\partial q}-\mathcal{V}'(q)\frac{\partial}{\partial v}\, ,\qquad
\bar{\Gamma}(\bar{q}, \bar{v})= \bar{v}\frac{\partial}{\partial \bar{q}}+\left(\frac{f'}{f}\bar{v}^2-f^{2}\mathcal{V}'(\bar q)\right)\frac{\partial}{\partial \bar{v}}\, .
$$  
But it is known that the energy function $E= \frac 12 v^2+\mathcal{V}= \frac 12( \bar v/f)^2+\mathcal{V}$ is a conserved quantity and
if we restrict ourselves to study the motions for a given energy $E$, 
   \begin{equation}
   \frac{d\bar v}{d\tau}=  f  f'(q)\,2(E-\mathcal{V})-f^2\, \mathcal{V}'(q)=\frac d{dq}\left(f^2(E-\mathcal{V})\right),\label{eqbarvV}
 \end{equation} 
 and then this is an inhomogeneous  linear differential equation in the variable $q$   iff there exist 
constants $A$, $B$ and $C$ such that $f^2(E-\mathcal{V})=A\,q^2+B\, q+C$, from where the final equation is 
  $$
  \frac{d\bar v}{d\tau}=  2\, A\, \bar q+B\Longrightarrow \frac{d^2\bar q}{d\tau^2}=2\, A\, \bar q+B.
  $$

In the very relevant  case of the radial equation for a given fixed  angular momentum $\ell$ for Coulomb--Kepler problem for which $q$ is the radial variable $r$ 
and $V(r)=-k/r$, we have 
$$
\mathcal{V}(r)=V(r)+\frac{\ell^2}{2r^2}=-\frac kr+\frac{\ell^2}{2r^2},\qquad  F(r)=-\mathcal{V}'(r)=\frac {\ell^2}{r^3}-\frac k{r^2},
$$
and from $f^2\, (E-\mathcal{V})=Ar^2+B\, +C$, we obtain
$$f^2\left(E+\frac kr-\frac{\ell^2}{2r^2}\right) = Ar^2+B\,r +C,
$$
i.e. 
$$
f^2\,E=A\, r^2,\quad f^2\, k= B\,r^2,\quad \ell^2\, f^2=-2\, C\, r^2,
$$
and therefore, the general solution is a multiple of $f(r)  = r$ with 
$$A= E,\quad B=k, \quad C=-\frac \ell 2,
$$
and then the transformed second-order differential equation is:
$$
\frac{d^2 r}{d\tau^2}=2\, E\, r+k,
$$
 which is the result given in \cite{KOP}. The remarkable fact is that the 
Sundman transformation  is independent of $\ell$,  because  $f(r)$ is proportional to $r$. 

 As another particular   example in the opposite direction, we can start from the  linear SODE vector field $\Gamma$ describing the time evolution of the 1-dimensional harmonic
  oscillator, i.e. $\Gamma= v\,(\partial/\partial x)- \omega^2 \, x\,(\partial/\partial v)$, which corresponds to $F(x)=-\omega^2 x$, i.e. $\mathcal{V}=\frac 12 \omega^2 x^2$. In this case,  under the Sundman transformation defined by a given  function $f$, 
 $$\bar{\Gamma}(\bar x,\bar v)=f(\bar x)\,\Gamma(x,\bar v)=\bar{v}\frac{\partial}{\partial \bar{x}}+
\left(\frac{1}{f}\frac{d f}{d \bar x}\bar{v}^2-f^2\, \omega^{2}\, {\bar x} \right)\frac{\partial}{\partial \bar{v}}\, ,\quad \bar v^i=f\, v^i,
 $$
whose integral curves are such that their projections on the base manifold are solutions of the differential equation 
  $$
  \frac{d^2\bar x}{d\tau ^2}- \frac 1{f}\frac{d f}{ d\bar x}\left(\frac {d\bar x}{d\tau}\right)^2+f^2\, \omega^{2}\, {\bar x}=0,
  $$
  i.e. it appears an additional  quadratic damping term,  and linearity of harmonic oscillator  is lost. Conversely, an equation of this last type can be reduced to a harmonic oscillator by means of the Sundman transformation defined by the function $f^{-1}$. If, for instance,  $f(x)=x^2$, 
 the former equation  reduces to 
 $$
 \frac{d^2\bar x}{d\tau ^2}+ \frac 2{\bar x} \left(\frac {d\bar x}{d\tau}\right)^2+ \omega^{2}\, \frac 1{\bar x^3}=0,
 $$
 i.e. it is the Ermakov-Pinney equation corresponding to  free motion under the action of a damping quadratic term \cite{RR80}. This is a prototype for linearisable examples to be analysed next.

\subsection{Generalised Sundman transformations}

\indent\indent  Let us remark that from a geometric viewpoint only Sundman transformations $dt=f(x)\, d\tau $ have a sense, i.e. the function $f$ does not depend on $t$, which is not a coordinate. 
However, as it was stated in Section~\ref{section:linear.sodes}, in adapted coordinates $y$, a linear structure on the base manifold $Q=\R$ is of the form $\Delta_Q=y\,\partial/\partial {y}$ and the linear character of a given  SODE depends on the choice of the coordinate $y$.  The strategy that we will follow in the study of the linearisation process  is to find a coordinate transformation $y=\varphi (x)$ from the original coordinate to the adapted one, and a Sundman transformation $d\tau =h(x)\, dt$ which transforms our original SODE into a linear one in the new coordinate $y$ and its velocity. This will be done in several steps, by finding several coordinate transformations and Sundman transformations which simplify the form of the SODE. In other words, we admit composition of ordinary changes of coordinates with the Sundman transformations considered until now. In this context we will refer to a transformation of the form $y=x$, $d\tau =h(x)\,dt$ as a pure Sundman transformation, and to $y=\varphi (x)$, $d\tau =dt$ as a pure coordinate transformation. By composition of such type of transformations we get a group of generalised Sundman transformations $(h,\varphi)$ defined as 
\begin{equation}
y=\varphi(x), \qquad d\tau=h(x)\ dt,\label{phihST}
\end{equation}
with composition law
$$(h_2,\varphi_2)\star (h_1,\varphi)=((h_2\circ \varphi_1) h_1, \varphi_2\circ \varphi_1).
$$
The neutral element is $(1,{\rm id})$ and the inverse of $(h,\varphi)$ is  $((1/h)\circ\varphi^{-1},\varphi^{-1}).$
The pure Sundman transformations are those of the form $(h, {\rm id})$, and close on an Abelian invariant subgroup,  while usual coordinate transformations correspond to those of the form 
$(1, \varphi)$ and made up also a subgroup. As each transformation can be factorised as 
$$ 
(h,\varphi)=(1,\varphi)\star (h,{\rm id})=(h\circ\varphi^{-1},{\rm id})\star (1,\varphi),
$$
the set of generalised Sundman transformations is a semidirect product group. 

\subsection{Linearisation under generalised Sundman transformations}

\indent \indent We will say  that a SODE $\ddot x=X(x,\dot x)$  is fibre-linearisable (or linearisable  in velocities) under generalised Sundman transformations  if it can be transformed to a SODE of the form $\ddot{x}+A(x)\dot{x}+b(x)=0$, where $A$ and $b$ are real functions, while  
we will say that the  SODE is linearisable under generalised Sundman transformations  if it can be transformed into a SODE of the form $\ddot{x}+\alpha\, \dot{x}+Bx+C=0$ for some real numbers $\alpha,B,C\in\R$.  Remark that when $C\ne 0$ the transformed equation is an inhomogeneous linear equation.

Consider a scalar SODE 
\begin{equation}\frac{d^2x}{dt^2}=X\left(x,\frac{dx}{dt}\right), \label{SODEq}\end{equation}
i.e. a generic autonomous second-order differential equation.  In order to study the linearisability of  such equation remark that the abovementioned group properties of the set of 
generalised Sundman transformations show that the possible linearising transformations will be factorisable as a composition of a coordinate transformation first and a pure 
Sundman transformation later, leading to the  linear equation. Inverting the process we can see first the form of the image under a pure Sundman transformation of the prototype  linear equation. This is given by the the general transformation rule (\ref{teSTsystem2}) for the 1-dimensional case, and then we see that the image is an equation of the  class of SODE{s} of the form 
\begin{equation}
\frac{d^2x}{dt^2}+\gamma _0(x)\left(\frac{dx}{dt}\right)^2+A_0(x)\frac{dx}{dt}+b_0(x)=0, \label{Dsv}
\end{equation}
i.e. the function $X(x,v)$ is a polynomial of degree at most two in the variable $v={dx}/{dt}$, or in other words, $\partial {^3X}/\partial {v^3}=0$.  But such a class is invariant under changes of coordinates because if we consider $\bar x=\varphi(x)$, then 
$$\frac{d\bar x}{dt}= \frac {d\varphi}{dx} \frac {dx}{dt},\qquad \frac{d^2\bar x}{dt^2}=\frac{d^2\varphi}{dx^2}\left( \frac {dx}{dt}\right)^2+ \frac {d\varphi}{dx} \frac {d^2x}{dt^2},
$$
and then $\bar x$ satisfies an equation of the same type than (\ref{Dsv}). Therefore we obtain as a first result that 
 a necessary condition for fibre-linearisability  under generalised Sundman transformations is that the function $X(x,v)$ be a polynomial of degree at most 2 in the variable $v={dx}/{dt}$, or in other words, $\partial {^3X}/\partial {v^3}=0$, and consequently 
 second-order differential equations that are  linearisable   under generalised Sundman transformations must be  of the form
 (\ref{Dsv}).
Such a class of second-order differential equations    is invariant under generalised  Sundman transformations and contains the subset of inhomogeneous linear in velocities equations, which correspond to
equations (\ref{Dsv})  with  $\gamma_0=0$. 
	 As an instance we can say that the example of  the Rayleigh-like oscillator studied in \cite{S20a},
 $$\ddot x+f(x)\,\dot x^3+g(x)\dot x^2+h(x)\dot x+k(x)=0,
$$
is not linearisable under generalised  Sundman transformations.

 We must determine which ones of these equations (\ref{Dsv}) are linearisable. 
The quadratic in the velocity  term of (\ref{Dsv}) can always be eliminated by an appropriate  pure Sundman transformation $x_1=x$, $dt_1=h(x)\, dt$. Indeed, under such a transformation the SODE takes the form
\[
\frac{d^2x_1}{dt_1^2}+\left(\gamma _0+\frac{h'}{h}\right)\left(\frac{dx_1}{dt_1}\right)^2+\frac{1}{h}A_0\frac{dx_1}{dt_1}+\frac{1}{h^2}b_0=0,
\]
so that the new coefficients are 
\begin{equation}
\gamma _1=\gamma _0+\frac{h'}{h},
\qquad
A_1=\frac{A_0}{h}
\qquad\text{and}\qquad
b_1=\frac{b_0}{h^2}.\label{ncuST}
\end{equation}
We can choose as function  $h=h_0(x)$ a non trivial solution of the linear \textsc{ode} 
\[
\frac{dh}{dx}+\gamma _0(x)h=0,
\]
whose  general solution is $h(x)=K\exp\left(-{\displaystyle \int} \gamma _0(x)\,dx\right)$ with $K$ being  any constant, which is irrelevant for our purposes.
 With this choice for the function $h$ the transformed second-order differential equation under the pure Sundman transformation defined by $h$ becomes the fibre-linear equation
\begin{equation}
\frac{d^2x_1}{dt_1^2}+A_1(x_1)\frac{dx_1}{dt_1}+b_1(x_1)=0,\label{fleq}
\end{equation}
with $A_1={A_0}/{h_0}$ and $b_1={b_0}/{h_0^2}$.
Therefore any SODE of the given class (\ref{Dsv})  is  fibre-linearisable by a pure Sundman transformation.

At this point we should notice that if  both $A_1$ and $b_1$ vanish identically (i.e. $A_0$ and $b_0$ both vanish identically) then we have already got a special linear equation, more specifically,
\[
\frac{d^2x_1}{dt_1^2}=0.
\]

Let us consider now the case where at least one of such functions is not the zero function. Remark that once we have got an inhomogeneous  fibre-linear SODE we can use only Sundman transformations that transform an inhomogeneous  fibre-linear SODE into a new inhomogeneous   fibre-linear SODE. These are slightly more general than affine transformations of coordinates, as it is stated in the following result:

\begin{lem}
{\it i)} If a generalised Sundman transformation  $\bar x=\varphi(x), \ d\bar{t}=h(x)\ dt$ preserves the set of inhomogoneous  fibre-linear SODEs,  then there exists a real  constant $c\neq 0$ such that
\begin{equation}
\frac{d\varphi}{dx}=c\,h.\label{lemacond}
\end{equation}

{\it ii)} A Sundman transformation  $\bar x=\varphi(x), \ d\bar{t}=h(x)\ dt$ such that (\ref{lemacond}) holds,  transforms every inhomogeneous fibre-linear SODE 
\begin{equation}
\frac{d^2x}{dt^2}+A(x)\frac{dx}{dt}+b(x)=0\label{eq1lem}
\end{equation}
into the  inhomogeneous  fibre-linear SODE 
\begin{equation}
\frac{d^2\bar{x}}{d\bar{t}^2}+\bar{A}(\bar{x}) \frac{d\bar x}{dt}+\bar{b}(\bar{x})=0,
\label{eq2lem}
\end{equation}
where
\begin{equation}
\bar{A}(\bar{x})=\frac{1}{h(x)}A(x)
\qquad\text{and}\qquad
\bar{b}(\bar{x})=\frac{c}{h(x)}b(x).
\label{eq3lem}
\end{equation}
\end{lem}
{\sl Proof.-}
{\it i)} Under a Sundman transformation $\bar{x}=\varphi (x)$, $d\bar{t}=h(x)\, dt$ we have 
\begin{equation}
\bar{v}=\frac{d\bar{x}}{d\bar{t}}=\frac{1}{h}\frac{d\varphi }{dx}\frac{dx}{dt}=\frac{1}{h}\frac{d\varphi }{dx}v,\label{barvv}
\end{equation}
and
\begin{equation}
\frac{d^2\bar{x}}{d\bar{t}^2}=\frac{1}{h}\left[\frac{d}{dx}\left(\frac{1}{h}\frac{d\varphi }{dx}\right)v^2+\left(\frac{1}{h}\frac{d\varphi }{dx}\right)\frac{d^2x}{dt^2}\right].\label{eq4lem}
\end{equation}
Therefore, if the original equation  is inhomogeneous fibre-linear as in (\ref{eq1lem}), then the transformed one  (\ref{eq4lem}) is also inhomogeneous  fibre-linear if and only the coefficient of $v^2$ vanishes, that is, if and only if the function $({1}/{h})({d\varphi }/{dx})$ is constant, which proves condition (\ref{lemacond}).

{\it ii)} If the Sundman transformation $\bar{x}=\varphi (x)$, $d\bar{t}=h(x)\, dt$ satisfies condition (\ref{lemacond}), we have that 
\[
\frac{d^2\bar{x}}{d\bar{t}^2}=\frac{c}{h(x)}\frac{d^2x}{dt^2},
\]
or in other words, it looks like the inhomogeneous fibre-linear equation   (\ref{eq2lem}) with
\[
\bar{A}(\bar{x})\bar{v}+\bar{b}(\bar{x})=\frac{c}{h(x)}\left(A(x)v+b(x)\right).
\]
Since in this case, according to (\ref{barvv}),  $\bar{v} =c\, v$ we find that
\[
\bar{A}(\bar{x})=\frac{1}{h(x)}A(x)
\qquad\text{and}\qquad
\bar{b}(\bar{x})=\frac{c}{h(x)}b(x),
\]
which ends the proof.

\qed

The explicit value of the constant $c$\,  does not have any influence in the problem of linearisability,  and then without loosing of generality we can  take $c=1$ (or $c=-1$ when appropriate). 

This  result shows that  we can only transform the inhomogeneous fibre-linear SODE (\ref{fleq})
 by means of a special kind of Sundman transformations $x_2=\varphi (x_1)$, $dt_2=h(x_1)\, dt_1$, with $\varphi (x)={\displaystyle \int^x} h(\zeta)\,d\zeta$, since, otherwise,  the transformed equation of (\ref{fleq}) would be non linear in the fibre variable. For such a transformation ${dx_2}/{dt_2}={dx_1}/{dt_1}$ and the expression of the transformed SODE is
\begin{equation}
\frac{d^2x_2}{dt_2^2}+\frac{1}{h(x_1)}A_1(x_1)\frac{dx_2}{dt_2}+\frac{1}{h(x_1)}b_1(x_1)=0,
\label{fleq2}
\end{equation}
where the substitution of $x_1$ by  $\varphi ^{-1}(x_2)$ on the last two terms is understood. This SODE is linear if, and only if, the function  $A_1/h$ is constant and  the function $b_1/h$ is affine in the variable~$x_2$.
 In the differential equation (\ref{fleq2})  we have  two different situations:
 
 a)  If $A_1$ vanishes identically (recall that also  $A_0$ vanishes identically),  we can choose the function $h$ as $h=|b_1|$ in the Sundman transformation and the 
 transformed SODE is then $$\frac{d^2x_2}{dt_2^2}+\beta =0,$$ with $\beta =\sign(b)$, which we assume to be constant (otherwise we have to restrict $x_2$  to an interval where the sign of $b$ is constant). The Sundman transformation is $x_2={\displaystyle\int^{x_1}}|b(\zeta)|\, d\zeta$ and  $dt_2=|b(x_1)|\, dt_1$. Alternatively, we can take the generalised Sundman transformation $x_2={\displaystyle\int^{x_1}} b(\zeta)\, d\zeta$, \, $dt_2=|b(x_1)|\,dt_1$, and the transformed SODE is 
\[
\frac{d^2x_2}{dt_2^2}+1=0.
\]

b)   If $A_1$ is not the zero function,  in order to make  constant the coefficient of $dx_2/dt_2$ we must take $h(x_1)=|A_1(x_1)|$ (up to an irrelevant multiplicative constant). Thus the transformation $x_2={\displaystyle \int^{x_1}} A_1(\zeta)\,d\zeta$, $dt_2=|A_1(x_1)|\, dt_1$,  transforms the given inhomogeneous fibre-linear SODE (\ref{fleq}) into the form
\begin{equation}
\frac{d^2x_2}{dt_2^2}+\alpha \frac{dx_2}{dt_2}+b_2(x_2)=0,
\label{flcceq}
\end{equation}
with $\alpha =\sign(A)$ (which once again we assume to be constant) and 
\[
b_2(x_2)=\frac{b_1(x_1)}{|A_1(x_1)|},
\]
where on the right-hand side we assume that $x_1$ is replaced by its corresponding value of $x_2$.

A generalised Sundman transformation that preserves the form of the above SODE (\ref{flcceq}), i.e.   inhomogeneous fibre-linear with constant coefficient $\alpha$,  is of the form $\bar{x}=m\,x_2+n$, $d\bar{t}=m\, dt_2$, with $m$ and $n$ some constants. Therefore, a SODE of the above form (\ref{flcceq}) is either linear or otherwise it is not linearisable. 
Obviously, the new SODE is linear if, and only if,  $b_2(x_2)$ is an affine function. In other words if, and only if, ${db_2}/{dx_2}$ is constant,
\begin{equation}
\frac{db_2}{dx_2}=B\in\mathbb{R}.
\label{Bconst}
\end{equation}

Let us find the conditions on the original data $\gamma _0$, $A_0$ and $b_0$ in order to $b_2(x_2)$ be an affine function. Using the chain rule and the definitions  (\ref{ncuST}) of $A_1$ and $b_1$, we see that 
\begin{align*}
B=\frac{dx_1}{dx_2}\frac{d}{dx_1}\left(\frac{b_1}{A_1}\right)
=\frac{1}{A_1}\frac{d}{dx}\left(\frac{b_0}{A_0h_0}\right) 
=\frac{1}{A_0^3}\left(A_0\frac{d}{dx}+\gamma A_0-A'_0\right)b_0.
\end{align*}
Taking into account that if $z$ is the real function 
$$z=\left(A_0\frac{d}{dx}+\gamma A_0-A'_0\right)b_0, $$
we have
\[
\frac{d}{dx}(A_0^{-3}z)=A_0^{-3}z'-3A_0^{-4}A'_0z=A_0^{-4}(A_0z'-3A'_0z),
\]
we get that the condition (\ref{Bconst}) can be rewritten as
\[
\frac{dB}{dx}=\frac{1}{A_0^4}\left(A_0\frac{d}{dx}-3A'_0\right)\left(A_0\frac{d}{dx}+\gamma _0A_0-A'_0\right)b_0=0.
\]
Consequently, we have arrived to the following linearisability condition:  the SODE (\ref{Dsv}) is linearisable if and only if the functions $A_0$, $b_0$ and $\gamma _0$ satisfy
\[
\left(A_0\frac{d}{dx}-3A'_0\right)\left(A_0\frac{d}{dx}+\gamma _0A_0-A'_0\right)b_0=0.
\]
Notice that this condition  is also satisfied in the two first cases (either $A_0=b_0=0$, or  $A_0\neq 0$ and $b_0=0$).  Moreover, we can prove that  such condition is   invariant under a generalised Sundman transformation:

\begin{theorem}
For the class of SODE{s} of the form 
\[
\frac{d^2x}{dt^2}+\gamma (x)\left(\frac{dx}{dt}\right)^2+A(x)\frac{dx}{dt}+b(x)=0,
\]
let $Q$ be the function 
\begin{equation}
\label{condition.Q=0}
Q= \left(A\frac{d}{dx}-3A'\right)\left(A\frac{d}{dx}+\gamma A-A'\right)b.
\end{equation}
Then, the condition $Q=0$ is invariant under generalized Sundman transformations.
\end{theorem}
{\sl Proof.-}
We have seen   that under a pure Sundman transformation, $d\tau =h(x)\, dt$, the coefficient functions transform as in (\ref{ncuST}):
\[
\gamma \mapsto\gamma +\frac{h'}{h},
\qquad
A\mapsto\frac{A}{h}
\qquad\text{and}\qquad
b\mapsto\frac{b}{h^2}.
\]
Thus the factor $$P=\left(A\frac{d}{dx}+\gamma A-A'\right)b$$ is transformed into 
$$
\left(\frac{A}{h}\frac{d}{dx}\!+\!\left(\gamma +\frac{h'}{h}\right)\!\frac{A}{h}\!-\!\left(\frac{A}{h}\right)'\right)\!\frac{b}{h^2}
\!=\!\left(\frac{A}{h}\frac{d}{dx}\!+\!\left(\gamma +\frac{h'}{h}\right)\!\frac{A}{h}\!-\!\frac{A'}{h}\!+\!\frac{Ah'}{h^2}\right)\!\frac{b}{h^2}
\!=\! \frac{1}{h^3}\!\left(\!A\frac{d}{dx}\!+\!\gamma A\!-\!A'\right)b,
$$
i.e. $P$ is transformed into $P/{h^3}$.

Hence $Q=\left(A\dfrac{d}{dx}-3A'\right)P$ is transformed into
$$
\left(\frac{A}{h}\frac{d}{dx}-3\left(\frac{A}{h}\right)'\right)\frac{P}{h^3}
=\left(\frac{A}{h}\frac{d}{dx}-3\frac{A'}{h}+3\frac{Ah'}{h^2}\right)\frac{P}{h^3}
=\frac{1}{h}\left(A\frac{d}{dx}-3A'+3\frac{h'}{h}A\right)\frac{P}{h^3},
$$
and using that 
$$
\frac d{dx}\left (\frac P{h^3}\right)=\frac 1{h^3}\ \frac d{dx} P-3\frac{h'}{h}P,
$$
we see that $Q$ is transformed into 
$$
\frac{1}{h^4}\left(A\frac{d}{dx}-3A'\right)P=\frac{1}{h^4} Q.
$$
Therefore $Q\mapsto\dfrac{1}{h^4}Q$, and hence the condition $Q=0$ is invariant under pure Sundman transformations.

Under a usual change of coordinates $x\mapsto \bar{x}=\varphi(x)$ the coefficient functions change as
\[
A\mapsto A
\qquad
b\mapsto J\,b
\qquad\text{and}\qquad
\gamma \mapsto \frac{1}{J}\left(\gamma -\frac{J'}{J}\right),
\]
where $J(x)={d\varphi}/{dx}$, and of course ${d}/{dx}\mapsto ({1}/{J})\,{d}/{dx}$, and hence $A'\mapsto ({1}/{J})\,A'$. The term $P=\left(A\dfrac{d}{dx}+\gamma A-A'\right)b$ is transformed into 
$$\begin{array}{rl}
{\displaystyle\left(A\frac{1}{J}\frac{d}{dx} \right.} &{\displaystyle+\left.\frac{1}{J}\left(\gamma -\frac{J'}{J}\right)A-\frac{1}{J}A'\right)(J\,b)
=\frac{1}{J}\left(A\frac{d}{dx}+\left(\gamma -\frac{J'}{J}\right)A-A'\right)(J\,b)}\\
&{\displaystyle=\left(A\frac{d}{dx}+\left(\gamma -\frac{J'}{J}\right)A-A'+A\frac{J'}{J}\right)b=\left(A\frac{d}{dx}+\gamma -A'\right)b},
\end{array}
$$
i.e. $P$ is invariant.

Hence $Q=\left(A\dfrac{d}{dx}-3A'\right)P$ is transformed into
\begin{align*}
\left(AJ\frac{d}{dx}-3JA'\right)P
&=J\left(A\frac{d}{dx}-3A'\right)P.
\end{align*}
Therefore,  $Q\mapsto JQ$, and we can conclude that the condition $Q=0$ is also invariant under pure coordinate transformations.

As a generalised  Sundman transformation can be obtained by composition of a pure coordinate transformation and a pure Sundman transformation,   the invariance of the condition $Q=0$ under generalised  Sundman transformations follows.

\qed

As in the particular case of  a linear SODE we have $Q=0$, this shows that the given condition is also necessary. Summarizing our results, we have proved the following statement.

 \begin{theorem}
\label{linearization}
i) A second-order differential equation $$\frac{d^2x}{dt^2}=X\left(x,\frac{dx}{dt}\right)$$ is fibre-linearisable by a Sundman transformation $y=\varphi (x)$, $d\tau =h(x)\, dt$,  if, and only if, it is of the form 
\begin{equation}
\label{quadratic}
\frac{d^2x}{dt^2}+\gamma (x)\left(\frac{dx}{dt}\right)^2+A(x)\frac{dx}{dt}+b(x)=0.
\end{equation}
Moreover, it can always be transformed to a constant coefficient $\alpha$ form 
\[
\frac{d^2y}{d\tau ^2}+\alpha \frac{dy}{d\tau }+\beta (y)=0,
\]
with $\alpha =\sign(A)$ (understanding that $\alpha =0$ if $A=0$).

ii) A second-order differential equation is linearisable if and only if it is of the form~\eqref{quadratic} and the coefficients $\gamma (x)$, $A(x)$ and $b(x)$ satisfy
\begin{equation}
\label{linearizability.condition}
Q\equiv\left(A\frac{d}{dx}-3A'\right)\left(A\frac{d}{dx}+\gamma A-A'\right)b=0.
\end{equation}

More specifically:
\begin{itemize}
\item If $A=0$ and $b=0$, then the SODE can be transformed into the form
\begin{equation}
\frac{d^2y}{d\tau ^2}=0\label{red1}
\end{equation} 
by the Sundman transformation 
\begin{equation} 
y=x, \qquad
d\tau =\exp\left(-\int^x \gamma (\zeta)\,d\zeta\right)\,dt.
\label{ST1}
\end{equation}

\item If $A=0$ and $b\neq 0$, then the SODE can be transformed into the form
\begin{equation}
\frac{d^2y}{d\tau ^2}+1=0\label{red2}
\end{equation} 
by the Sundman transformation 
\begin{equation}y=\int^x b(\zeta)\exp\left(2\int^\zeta \gamma (\eta)\,d\eta\right)\,d\zeta, \qquad 
d\tau =|b(x)|\exp\left(\int^x \gamma (\zeta)\,d\zeta\right)\,dt.
 \label{ST2}
\end{equation}

\item If $A\neq 0$ and condition~\eqref{linearizability.condition} is satisfied, then the SODE can be transformed into the form 
\begin{equation}
\frac{d^2y}{d\tau ^2}+\alpha \frac{dy}{d\tau }+By+C=0,\label{red3}
\end{equation} 
where $\alpha =\sign(A)$, by the Sundman transformation
\begin{equation}
y=\int^x A(\zeta)\exp\left(\int^\zeta \gamma (\eta)\,d\eta\right)\,d\zeta, \qquad
d\tau =|A(x)|\, dt.
\label{ST3}
\end{equation}
\end{itemize}
 \end{theorem}
 
  $\hfill\Box$
 
 \section{Some examples of linearisable under Sundman transformations systems}
 
 In this section we illustrate the theory of linearisable systems with some particular examples:
  
\begin{example}[Ermakov-Pinney~\cite{RR80}]
Consider once again  the differential equation
\[
\ddot{x}+\frac{2}{x}\dot{x}^2+\frac{\omega ^2}{x^3}=0,\qquad \omega\in\mathbb{R},
\]
so that $\gamma (x)=2/x$, $A(x)=0$ and $b(x)=\omega ^2/x^3$. Therefore it is Sundman linearisable and can be transformed to the form (\ref{red2}) by means of the transformation 
given by (\ref{ST2})
\[
y=\omega ^2x^2,
\qquad
d\tau =\frac{\omega ^2}{x}\, dt.
\]
\end{example}

\begin{example}[Geodesics on the sphere~\cite{HARRI02}]
Consider as a fully analogous example the differential equation
\[
\ddot{x}=2\dot{x}^2\cot x+\sin x\cos x,
\]
so that $\gamma (x)=-2\cot x$, $A(x)=0$ and $\,b(x)=-\sin x\cos x$. It is Sundman linearisable by means of the transformation  given by (\ref{ST2})
\[
y=\frac{1}{2\sin^2(x)},
\qquad
d\tau = |\cot(x)|\ dt,
\]
and the transformed SODE is once again (\ref{red2}).
\end{example}

\begin{example}
Consider the differential equation~\cite{NAP10} 
\begin{equation}
\ddot{x}+\frac{1}{x}\dot{x}^2+x\, \dot{x}+\frac{1}{2}=0,\label{eqNAP}
\end{equation}
so that $\gamma (x)=1/x$, $A(x)=x$ and $\,b(x)=1/2$. The condition $Q=0$ is trivially satisfied because
\[
\left(A\frac{d}{dx}+\gamma A-A'\right)b=\left(x\frac{d}{dx}\right)(1/2)=0,
\]
and therefore it is Sundman linearisable. If we consider the interval $x>0$ then the Sundman transformation  given by (\ref{ST3})
\[
y=\frac{1}{3}x^3,
\qquad
d\tau=x\, dt
\]
transforms the given SODE into  the form given by (\ref{red3}),
\[
\frac{d^2y}{d\tau^2}+\frac{dy}{d\tau}+\frac{1}{2}=0.
\]
Notice that this equation (\ref{eqNAP}) is not linearisable by point transformations, as it does not satisfy Lie criteria, but it is Sundman linearisable.
\end{example}

\begin{example}
As a generalization of the  preceding example, we consider the differential equation
\[
\ddot{x}+\frac{1}{x}\dot{x}^2+x\,\dot{x}+b(x)=0.
\]
The condition $Q=0$ is  
\[
\left(A\frac{d}{dx}-3A'\right)\left(A\frac{d}{dx}+\gamma\, A-A'\right)b=(\Delta_\R-3)\Delta_\R b=0,
\]
with $\Delta_\R=x\partial/\partial {x}$. Therefore $Q=0$ if and only if  $\Delta_\R b$ is homogeneous with degree $3$, say $\Delta_\R b= 3k_1x^3$ with $k_1\in\R$, and hence $b(x)=k_1x^3+k_2$. It follows 
that the more general SODE of the above form that is Sundman linearisable is $\ddot{x}+\dfrac{1}{x}\dot{x}^2+x\dot{x}+k_1x^3+k_2=0$. A Sundman transformation  linearising this equation  is $y=\frac{1}{3}x^3$, $d\tau=|x|\, dt$.
\end{example}
 
\begin{example} 
 The Li\'enard equation 
 \begin{equation}
 \ddot x+f(x)\,\dot x+g(x)=0,\label{lieneq}
  \end{equation}
 and the existence of generalised Sundman transformations 
 $$y=\varphi(x) ,\quad  d\tau= F(x)\ dt,
 $$
able  to transform the given equation into 
 $$\frac {d^2y}{d\tau^2}+3\frac{dy}{d\tau}+ y^3+2y=0,$$
  are studied in  \cite{KS16}.
 
 Moreover, the generalised Sundman   transformation leading from the original equation to the linear equation 
 $$\frac {d^2y}{d\tau^2}+\sigma\frac{dy}{d\tau}+y=0$$
is also  found. 

Note that (\ref{lieneq}) is a particular case of the master equation
\begin{equation}
\frac{d^2x}{dt^2}+\gamma (x)\Bigl(\frac{dx}{dt}\Bigr)^2+A(x)\frac{dx}{dt}+b(x)=0,\label{mastereq}
  \end{equation}
 with $ \gamma(x)=0$, $A(x)=f(x)$  and $ b(x)=g(x)$. Recall also that the linearisability condition is 
 \begin{equation}
\label{linearisabilitycondition}
Q\equiv\Bigl(A\frac{d}{dx}-3A'\Bigr)\Bigl(A\frac{d}{dx}+\gamma A-A'\Bigr)b=0,
\end{equation}
that in our particular case turns out to be
$$
\left(f(x)\frac d{dx}-3 f'(x)\right) \left(f(x)\frac d{dx}- f'(x)\right) g=0,
$$
and therefore 
$$f^2\,g''-f\, f''\,g-3f\,f'\, g'+3f^{\prime 2}g=0.
$$
Then, given  the function $f$,  the function $g$ is any function of the linear space of solutions of the  linear   second-order differential equation in the variable $g$. But as $g=f$ is a solution of 
such equation we can introduce the change of variable $g=f\,\zeta$ and the given equation becomes 
$f\,\zeta''-f'\,\zeta'=0$, which shows that the general solution is
$$\zeta= k_1\int^x_0 f( \xi)\, d\xi+k_2,
$$
and therefore the linearisability condition implies that the function $g$ is
$$g(x)= k_1\, f(x)\int^x_0 f( \xi)\, d\xi+k_2\,f(x),
$$
in agreement with the result of Theorem 2 of \cite{KS16}. 

The example may be used to study the  Li\'enard type equation containing a dissipative term.
As explained in \cite{KS17}   the differential equation 
\begin{equation}
 \ddot x+f(x)\,\dot x^2+g(x)\,\dot x+h(x)=0\label{lieneq2}
  \end{equation}
  can be reduced by a pure Sundman transformation   
  $d\tau=F(x)\, dt$ to a Li\'enard equation (\ref{lieneq}). In fact, if 
  $$F(x)=\exp\left(-\int^xf(\xi\, d\xi\right)$$
  (\ref{lieneq2}) becomes
   \begin{equation}
   \frac{d^2x}{d\tau^2}+\tilde g(x)\frac {dx}{d\tau}+\tilde h(x)=0,\label{lieneqt}
    \end{equation}
  with
  $$\tilde g(x)= g(x)\exp\left(\int^xf(\xi\, d\xi\right), \qquad \tilde h(x)=h(x)\exp\left(2\int^xf(\xi\, d\xi\right).
  $$
  This process is the one indicated for going from  (\ref{Dsv})  to (\ref{fleq}) and shows that the linearisability under Sundman transformations of (\ref{lieneq2}) is reduced to 
  that of the corresponding equation (\ref{lieneqt}).

\end{example}

 \section{Conclusions and outlook}

 The geometric approach to Sundman transformation  defined by basic functions   for systems of second-order differential equations
 has been developed. It has been shown that, as  it also happens for systems of 
 first-order differential equations,  it amounts to replace the dynamical vector field by the corresponding conformally related one, but the additional price to be payed is the change of the usual tangent bundle structure by a new one that depends on the  basic function defining the Sundman transformation in such a way that the transformed dynamical vector field is a 
 SODE vector field with respect to the new tangent structure. The study is based on the use of quasi-coordinates on the tangent bundle that turn out to be true tangent bundle coordinates with respect to the new tangent structure. As an application we have developed the study of the linearisation of a second-order differential equation where not only standard 
Sundman transformation but a generalisation in which a  change of coordinates is also involved, because linearity depends on the choice of coordinates. Finally the  theory has been illustrated with several examples. 

Systems of  second-order differential equations   equivalent to a Euler-Lagrange system of second-order differential equations are  a privileged class because of its interest in many physical problems, and consequently, this particular class of systems is worth of a deeper study. Particularly interesting  are those systems describing geodesic motions in Riemann manifolds $(M,g)$ and the more general class 
of natural systems, also called of mechanical type,  where forces derivable of a potential function $V$ appear. This leads to study both geodesic motions
 for conformally related metrics \cite{TPBC13} and similar problems when potential functions are involved. 
These questions have been shown to be relevant in the study of classical superintegrable systems (see \cite{CRS21a,CRS21b} and references therein).  

\end{document}